% Standard macros that should be read at start of most of IHH's files.

\font\bigbf=cmbx10 scaled \magstep1 % Large Boldface
 % larger Bold
% Title macro. Note the \par at end to enforce the processing inside.
\def\title#1{\bgroup\leftskip 0 pt plus1fill \rightskip 0 pt plus1fill
\pretolerance=100000 \lefthyphenmin=20 \righthyphenmin=20
\noindent
#1
\par\egroup}

%\magnification=\magstep1
\baselineskip= 14pt plus .3pt minus .0pt    % comfortable spacing
% \parskip 5pt
% \headline{\hss Page \folio}  %  A headline

% Use the following two macros only in math mode.
\def\grapprox{\ \raise .6ex\hbox{$>$} \hskip -.11in\lower .5ex\hbox{$\sim$}\ }
\def\lsapprox{\ \raise .6ex\hbox{$<$} \hskip -.11in\lower .5ex\hbox{$\sim$}\ }

% Definitions to get bold symbols, e.g.Greek, and italics in equations.
% Use \bmath{}.  

%BOLD MATH FOR SINGLE CHARACTER

\newcount\yr
\yr=\year
\advance\yr -1900

\def\today{\number\day\ \ifcase\month\or
  Jan\or Feb\or Mar\or Apr\or May\or Jun\or
  Jul\or Aug\or Sep\or Oct\or Nov\or Dec\fi
  \space \number\yr}

% For input of HP laserjet figures. Old version for arbortex.
%\def\insertplot#1#2#3{\par
%  \hbox{%
%    \hskip #3                % Left margin.
%    \vbox to #2{             % Height.
%      \special{hp: plotfile #1}
%      \vfil
%    }%
%  }
%}
% For input of PCX figures. EMTEX Drivers. Use PCX file that matches 
% the plot size and has 118 or 300 resolution appropriately.
% Typical usage: \insertplot{file}{4 truein}{0.5 truein}

% This is to get around the hyphenation problem
\hyphenation{Mass-a-chu-setts In-sti-tute Tech-no-logy}
%\phantom{}  % get into vmode so that the first page can be started with vskip

\settabs 5\columns % a convenient general purpose table layout. (Use \+ &etc)

\magnification \magstep1
%\headline{Draft\hss\today}

\centerline{\bigbf Experimental Divertor Similarity Database Parameters}
\bigskip
\centerline{I.H.Hutchinson, B.LaBombard and B.Lipschultz}
\bigskip
\centerline{Plasma Fusion Center, MIT, Cambridge MA, USA.}
\bigskip
\beginsection{Abstract}

A set of experimentally-determined dimensionless parameters is
proposed for characterizing the regime of divertor operation. The
objective is to be able to compare as unambiguously as possible the
operation of different divertors and to understand what physical
similarities and differences they represent. Examples from Alcator C-Mod
are given.

\beginsection{Background}

Recent theory has shown that exactly scaled similarity experiments
on tokamak divertors, where atomic physics is important in addition to
plasma physics processes, are not feasible because there are no free
parameters. However, similarity comparisons can be obtained in
principle if either some of the plasma parameters [1], or some of
the geometric parameters [2], or both, are unimportant. Moreover, such
comparisons can be made practical if only approximate similarity is
required. Approaches to obtaining similarity with ITER divertor
regimes have been explored under various assumptions about divertor
transport scaling.

	There are now several moderately similar divertor tokamaks,
with increasing quantities of divertor data, that span a range of
dimensionless regimes. In addition to the theoretical approaches to
identifying similarity scaling, it seems most natural now to examine
the experimental results to discover what the values of relevant
dimensionless parameters actually are, and therefore to try to compare
from experiment to experiment so as to discover the underlying physics
that is similar or different in them. To do this requires a set of
dimensionless parameters to be indentified and, as far as possible,
measured on each experiment. With these stored in appropriate
databases, it ought to be possible quickly to identify dimensionless
regimes. Alternatively, the parameters can be evaluated from a less
processed database of raw data once the form of the parameters are
agreed, provided that the database contains the relevant raw data.

	The purpose of this note is to propose parameters that
determine the regimes of the relevant physics, {\it and} are
experimentally measurable. Since judgement must be exercised in the
choice of the exact parameters to include in the set, these choices
should not be regarded either as definitive or as exhaustive. However,
it is hoped that the set will form the basis for discussion, initial
comparisons, and further refinement.

\beginsection{The Formal Parameters}

As has been discussed elsewhere [1,2] the dimensionless parameters
deriving from the intensive variables (i.e.~density, temperature,
magnetic field, etc.) of the plasma can be considered
to be 

\item{$\rho^*$:} The ratio of Larmor radius to characteristic size.

\item{$\nu^*$:} The collisionality -- ratio of size to mean-free-path.

\item{$\lambda_D^*$:} The ratio of Debye length to size.

\item{$\beta$:} The ratio of plasma kinetic to magnetic pressure.

\item{$T^*$:} The ratio of plasma temperature to characteristic atomic energy.

\item{$f_j$:} The ion (or neutral) density of a species, $j$, as a
fraction of electron density.

\noindent Of these, it is commonly supposed that $\lambda_D^*$ is
unimportant. The parameters $f_j$ are undoubtedly important but often
not explicitly considered.

In addition to these parameters there is a host of other
dimensionless parameters that can be considered to be geometric.
These include

\item{$\bullet$} Relative shapes of the plasma and divertor cross-section,
including such quantities as divertor flux expansion factors and
relative depth.

\item{$\bullet$} Ratios of poloidal to toroidal field, and hence field-line
angle, safety factor, and so on.

\noindent Moreover, geometric considerations are essential for the
intensive parameters too, because their values depend upon the choice
of their exact definition and where they are measured. Therefore it
is important to be explicit about how the intensive parameters are to
be defined.

The parameters listed above are sufficient to define the similarity of
equivalent systems. If two systems are found with all these parameters
the same, then all relevant parameters are equal for the two systems.
However, in comparing systems which are not exactly similar, there are
other dimensionless parameters that describe important processes in
the divertor. It may be of interest to document their values
explicitly even if they are in principle determined by the parameters
above. We here propose one such vital parameter: the ratio of the
charge-exchange mean-free-path to the SOL thickness.

The following proposals for specific choices of parameters'
definitions are based on judgement about what is important and what it
is generally measured. They are therefore open to refinement. However,
it is hoped that some degree of agreement on parameter definitions can
be obtained along these lines.

\beginsection{The Specific Parameters}

There are, roughly speaking, three poloidal positions in the SOL that
may be considered characteristic. They are, (1) the main chamber, (2)
the x-point, (3) the target. By the ``main chamber'', we mean the part
of the SOL that is adjacent to the core plasma and sufficiently far
from the x-point as not to be significantly influenced by it. The
x-point is the transition between regions of the SOL adjacent to the
main plasma and to the private flux region. This point is
representative of the upstream divertor.  The target is where the
field lines intersect the solid surface. 

It is proposed that intensive dimensionless parameters be defined at
each of the three positions. Generally, in present experiments there
are measurements at the target and in the main chamber but perhaps not
at the x-point, although efforts are currently devoted to obtaining
this data.

 The radial scale-length with which plasma characteristic lengths are
to be compared is the width of the scrape-off layer, $\Delta$. It is
proposed that this be defined as half the pressure e-folding length.
This e-folding length is the distance from the separatrix
to where the plasma electron pressure ($nT$) is $\exp(-1)$ of its
value at the separatrix, or, equivalently for an exponential profile,
the ratio of the radial integral of the pressure to its separatrix
value.  The practical advantage of using the pressure is that, except
for detached divertors, the pressure is approximately constant along
the flux surface so the $\Delta$ can be evaluated either from
main-chamber or from target measurements. (A potential disadvantage is
that we need both $T$ and $n$ to be measured.) Half the exponential
width is used so as to be typical of the near-separatrix region. 
Note, of course, that $\Delta$ varies with poloidal position.

The flux surface at distance $\Delta$ is proposed also as the radial
position at which the plasma parameters such as $T_e$ and $n_e$ are to
be evaluated. Their values there will be slightly smaller (by about
20\%) than the separatrix values in the main chamber, which are most
often quoted.  However, the advantage of using the values at $\Delta$
is that they are more representative of the whole SOL. It is found
that the divertor target values exactly on the separatrix are very
often strongly influenced by the private flux region and by neutrals,
rendering them even less typical of the divertor fan as a whole.

For the gyrosize parameter, $\rho^*$, it is proposed that this be
defined for singly charged ions as
 $$
	\rho^* \equiv { (m_i T_e)^{1/2} \over e B_t} / \Delta \quad.
 $$
 This uses the ion Larmor radius at the sound speed, ignoring the ion
temperature. Corrections for finite $T_i$ can be incorporated as an
auxiliary parameter $T_i/T_e$. It also uses the toroidal field whose
variation as a function of position is weak, so that the $\rho^*$
variation from place to plase is caused primarily by temperature
differences.  Physically, the poloidal Larmor radius may be key. It
then requires the extra geometric parameter of the field-line angle,
$B_p/B_t$. The ratio of the local poloidal Larmor radius to the local
scrape-off length is almost independent of poloidal position (because
each has a $B_t/B_p$ factor and they cancel) and is equal to $\rho^*
B_t/B_p$ using midplane values.

The Coulomb collisionality of greatest interest is that for parallel
motions. It is proposed that the parallel connection length, $L$, for
the magnetic field be documented for two regions: the main chamber and
the divertor. We denote by $L_m$ the complete connection length from
target to target and by $L_x$ the connection length from x-point to
target. (The inner and outer divertor connection lengths may be
different in single-null configurations, and there are two different
total connection lengths for double-null configurations.) The
connection length has a logarithmic singularity at the separatrix
arising from the x-point region.  Connection lengths are therefore
measured on the flux surface that is a distance $\Delta$ outside the
separatrix, where the plasma parameters are measured. It is proposed
that the mean-free-path for Coulomb collisions be taken as that for
electron-ion momentum transfer, sometimes called ``slowing down''.
Using the plasma formulary [3] expression for thermal electrons,
$\nu_{e} = 2.9\times 10^{-6} (n/cm^{-3}) \ln \Lambda (T_e/eV)^{-3/2}$
and the characteristic speed, $(T_e/m_e)^{1/2}$ the electron
mean-free-path is

 $$
 \ell_e = 1.45\times 10^{-4} {(T_e/eV)^2 \over (n/10^{20}m^{-3})}
\quad m,
 $$
 taking the coulomb logarithm, $\ln \Lambda$, to be 10. We then define 
 $$
	\nu^* = L/\ell_e \quad,
 $$
 for the different connection lengths.

  It should be noted that the ion-ion mean-free-path is essentially
the same as the electron, at the same temperature. The ion-electron
mean-free-path is $(m_i/m_e)^{1/2}$ longer. It should also be noted
that the mean free path for epithermal electrons ($v\approx 3.7 v_t$)
responsible for heat transfer is much longer than the thermal value,
by a factor approximately 17 accounting for both electron and ion
collisions and using the fast electron expressions. This latter
collisionality determines whether the SOL is in the sheath- or
conduction-limited regimes.

 For the dimensionless temperature, relative to characteristic atomic
physics energy scale, there is no obvious atomic energy that ought to
be prefered; therefore the temperature in some convenient units is the
natural choice of definition. Electron volts seems the natural choice.
More importantly, the electron and ion temperatures may well be
different. For the atomic processes the electron temperature is
undoubtedly more important. The ion temperature is not completely
insignificant, expecially for such processes as charge exchange, but
its importance as a dimensionless parameter relative to atomic
energies is much less.

The key geometric parameters are associated with the plasma shape and
the field line angle. From the point of view of the scrape-off layer,
the significant shape parameters are presumably (1) the ratio of the
connection length in the divertor to the connection length in the main
chamber and (2) the ratio of the flux surface spacing in the divertor
to the spacing in the main chamber (i.e. the flux expansion factor),
which we will choose to define at a distance $\Delta$ outside the
separatrix so as to be well defined at the x-point poloidal position.
These together also determine the ratio of the divertor poloidal depth
to the plasma radius. Moreover, the flux expansion factor is equal to
the ratio of the field line angles in the divertor and main chamber.
However, one value of field line angle is still required to define the
field geometry, it can be taken simply as $B_p/B_t$ at the midplane.

A parameter that may also be important for the details of recycling
and related processes at the target is the angle in the poloidal plane
of intersection between the separatrix and the divertor target.  We
will denote this $\alpha_p$. A ``horizontal'' target plate generally
has $\alpha_p < \pi/2$, while a ``vertical'' plate has $\alpha_p >
\pi/2$. The total three-dimensional angle of intersection of the
field-line and the target ($\alpha_t$ say) is then given by
$\sin\alpha_t = \sin\alpha_p /(1 + B_t^2/B_p^2)^{1/2}$, evaluated at the
target.

An auxiliary parameter of great importance is the collisionality of
neutral flow across the divertor. For symmetric charge-exchange of
hydrogen, the reaction rate is ${\tt <}\sigma v{\tt >} \approx
10^{-14} T_i^{0.33}$ m$^{-3}$s$^{-1}$ for temperatures of interest.
Therefore, the mean-free-path can be taken as $\ell_{cx} = (2{\cal
E}/m)^{1/2}/n{\tt <}\sigma v{\tt >}$ m, where ${\cal E}$ is the neutral
energy.  We can consider specified energy (e.g.~cold or Franck-Condon)
neutrals, or else hot thermal neutrals with ${\cal E} = T_i/2$. We
define representative cases as the cold expression for 3 eV atoms:
 $$
 \ell_{\rm cx,cold} \approx 2.4\times10^{-2} [(T_i/eV)/\mu]^{-0.33}
(n_i/10^{20}m^{-3})^{-1}\quad m
 $$
and the hot expression for thermalized atoms:
 $$
 \ell_{\rm cx,hot} \approx 1.4\times10^{-2}  [(T_i/eV)/\mu]^{0.17}
(n_i/10^{20}m^{-3})^{-1}\quad m.
 $$
 Electron excitation and ionization processes are also of great
importance. However, the plethora of different rates makes the task of
defining explicit excitation lengths overwhelming. In order of
magnitude, the ionization rate above about 10 eV temperature is about
a factor of two less than the charge-exchange rate. So that the
ionization length is similar to the charge-exchange length. However,
as the electron temperature falls below 10 eV, ionization rapidly
becomes negligible.

We will now present expressions for the dimensional and dimensionless
parameters in table form. We use slightly abbreviated notation for
the dimensional parameters as noted in Table 1. Also $\mu$ denotes
the bulk ion mass in units of the proton mass.

\bigskip
\vbox{
\noindent Table 1. Dimensional Parameters.\par
\medskip
\line{\vrule\vbox{
\halign{\hss #\quad & # \quad & # \hss \cr
 \multispan{3}\hrulefill\cr
 \quad Parameter & Units & Definition \cr
 \multispan{3}\hrulefill\cr
 $n_{20}$ & ${\rm 10^{20} m^{-3}}$ & Electron density at $\Delta$-surface\cr
 $T$ & eV & Electron temperature at $\Delta$-surface\cr
 $B_t$ & T & Toroidal magnetic field\cr
 \multispan{3}\hrulefill\cr
 $L$ & m & Parallel connection length\cr
 $\Delta$ & m & Half radial midplane distance to 1/e pressure point.\cr
 \multispan{3}\hrulefill\cr
}}\vrule\hss
}
\noindent Qualified by position subscripts: $m$ main chamber, $x$
x-point, and $t$ target.
}

\par\bigskip
\vbox{
\noindent Table 2. Dimensionless parameters in practical units (eV,
m, T).\par
\medskip
\line{\vrule\vbox{
 \halign{\quad\hss #\quad & # \quad & # \hss \cr
 \multispan{3}\hrulefill\cr
 Parameter & Formula & Description\cr
 \multispan{3}\hrulefill\cr
 $L_x/L_m$ & & Relative parallel length of divertor.\cr
 $B_{pt}/B_{pm}$ & & Flux expansion factor at target.\cr
 $B_{px}/B_{pm}$ & & Flux expansion factor at x-point.\cr
 $B_{pm}/B_{tm}$ & & Field line angle in main chamber (midplane). \cr
 $\alpha_p$ & & Poloidal angle of intersection with target. \cr
 \multispan{3}\hrulefill\cr
 $\rho^*$ & $1.02\times10^{-4}{(\mu T)^{1/2} \over B_t \Delta}$ &
 Inverse of gyrowidth of SOL.\cr
 $\nu^*$ & $6.9\times10^{3}{L n_{20} \over T^2}$ & Parallel 
  Coulomb Collisionality. \cr
 $\beta$ & $8.0\times10^{-5}{n_{20}T \over B_t^2} $  & SOL toroidal beta. \cr
 $T$ & & Temperature\cr
 $T_i/T_e$ & & Relative ion temperature.\cr
 $f_j$ & $n_j/n_e$ & Relative species density.\cr
 \multispan{3}\hrulefill\cr
 $\Delta/\ell_{\rm cx,cold}$ & ${42 \Delta n_{20} (T_i/\mu)^{0.33} 
 %(T_i/T_e)^{.33}
 }$ & Transverse cold neutral collisionality.\cr
 $\Delta/\ell_{\rm cx,hot}$ &  ${71 \Delta  n_{20} (T_i/\mu)^{-0.17}
 %(T_i/T_e)^{-.17}
 }$ & Transverse thermalized neutral collisionality.\cr
 \multispan{3}\hrulefill\cr
}}\vrule\hss
}
\noindent Non-geometric quantities qualified by position
subscripts: $m$ main chamber, $x$ x-point, and $t$ target.
}

A spreadsheet [4] has been set up whereby the dimensionless parameters are
calculated once the dimensional parameters of Table 1 and additional
independent parameters (mostly geometric) of Table 2 are input.
An illustrative example of this spreadsheet is given in Table 3.

\beginsection{Alcator C-Mod Examples}

The parameters defined above have been included in the Alcator C-Mod
edge database and are calculated automatically, based primarily on
probe data (embedded and fast-scanning probes). We illustrate the
range of parameters on C-Mod by several figures based on operation
during March through April 1995. Fig 1 shows a plot of poloidal field
against separatrix density; the current grouping shows that the field
depends primarily on plasma current, as expected. (This data is all at
toroidal field of 5.3T.) Fig 2 plots the dimensionless parameters
$\rho^*$ versus $\nu^*$. The value of $\nu^*$ ranges from about 10 to
over 400, corresponding to the observed range of regimes from
sheath-limited to conduction-limited. Fig 3 plots $\beta^*$ versus
$\Delta$. The betas are low; the widths a few millimeters. All these
data are main-chamber parameters (denoted `mid' in the plots).

In Fig 4 we show the total connection length (target to target) as a
function of the ratio of $\Delta$ to the poloidal Larmor radius (i.e.
$(B_p/B_t)1/\rho^*$). Connection length varies primarily with current.
The SOL width is typically only a few times the poloidal gyrolength,
thus emphasizing the importance of cross-field drifts, as has been
experimentally observed. Fig 5 illustrates the values of the `cold'
neutral collisionality at the outer divertor (x-axis) and the main
chamber (y-axis). The neutral collisionality is generally less than
about 0.4 in the main chamber, implying that a significant fraction of
neutrals can cross the SOL.  At the divertor, however, the
collisionality is much higher (particularly at high densities) and the
divertor leg is therefore `thick' to neutrals.

These illustrative examples document the parameter ranges spanned by
this set of experiments. It is intended that inter-machine
comparisons be carried out using the definitions described here.

\parindent 0 pt
\beginsection{References}

[1] K.Lackner, Comments on Plasma Physics and Controlled Fusion, {\bf
15}, 359 (1994).

[2] I.H.Hutchinson and G.C.Vlases, to be published in Nuclear Fusion.

[3] D.L.Book, {\it Plasma Formulary}, NRL publication 0084-4040 (1986).

[4] This can be obtained in by anonymous FTP from
 //pfc.mit.edu/hutch/divsim.zip

\beginsection{Figure Captions}

Figure 1 Plot of similarity database parameters (SIM\_) poloidal
field versus separatrix density (at midplane).

Figure 2 Plot of $\rho^*$ versus $\nu^*$ (at midplane).

Figure 3 Plot of $\beta^*$ versus $\Delta$.

Figure 4 Plot of total connection length versus $\Delta/\rho_p$ (at
midplane). 

Figure 5 Plot of cold neutral transverse collisionality
($\Delta/\ell_{cx,cold}$)  at the midplane versus at the outer
divertor.

\vfill\eject

%%%%%%%%%%%%%%%%%%%%%%%%%%%%%%%%%%%%%%%%%%%%%%%%%%
%Mofigied from iopverb aug 94.
%
% Macros required for verbatim listings in text and as displays
% From The TeXbook and TeX for the Impatient
%
\font\tt=cmtt9
\chardef\other=12
\def\deactivate{%
  \catcode`\\=\other \catcode`\{=\other
  \catcode`\}=\other \catcode`\$=\other
  \catcode`\&=\other \catcode`\#=\other
  \catcode`\%=\other \catcode`\~=\other
  \catcode`\^=\other \catcode`\_=\other}

{\obeylines\gdef\startdisplay#1
  {\catcode`\^^M=5$$#1\halign\bgroup\indent##\hfil&&\qquad##\hfil\cr}}
\outer\def\enddisplay{\crcr\egroup$$}
\def\ttverbatim{\begingroup\deactivate\obeyspaces\obeylines \tt}
{\obeyspaces\gdef {\ }}  % \obeyspaces now gives \ , not \space
\catcode`\|=\active
{\obeylines
\gdef|{\ttverbatim\spaceskip=.5em plus.05em minus.05em%
\let^^M=\ \let|=\endgroup}}%

%
% \makeactive, \newline and \removebox together 
% make every line a paragraph
%
\def\makeactive#1{\catcode`#1=\active\ignorespaces}
{% The group delimits the text over which ^^M is active.
  \makeactive\^^M %
  \gdef\obeywhitespace{%
  % Use \gdef so the definition survives the group.
    \makeactive\^^M %
    \let^^M=\newline %
    \aftergroup\removebox % Kill extra paragraph at end.
    \obeyspaces %  
  }%
}
\def\newline{\par\noindent}
\def\removebox{\setbox0=\lastbox}
\def\vb{\goodbreak\vskip\abovedisplayskip%
\par\begingroup\tt % small typewriter font
\deactivate\obeywhitespace\catcode`\|=0 % Make | the new escape character.
}
\def\|{|}

%
%
% end iopverb.tex

\nopagenumbers
\baselineskip 11pt
\vb

    Table 3. Spreadsheet for Calculation of
    Experimental Divertor Similarity Parameters
                           Example
    Parameter  Units         Value
    B_t        T        * 4.00E+00
    L_m        m        * 2.20E+01
    L_x/L_m             * 1.60E-01
    B_pt/B_pm           * 3.70E-01
    B_px/B_pm           * 1.40E-01
    B_pm/B_tm           * 1.70E-01
    alpha_p             * 5.00E-01
    m_i/m_proton        * 2.00E+00
|medskip
    Main chamber
    n_20       10^20    * 8.40E-01
    T          eV       * 6.00E+01
    T_i/T_e             * 1.00E+00
    Delta      m        * 1.60E-03
    rho*                  1.78E-01
    nu*                   3.54E+01
    beta                  2.52E-04
    D/l_cold              2.18E-01
    D/l_hot               4.76E-02
    D/rho_p               9.55E-01
|medskip
    X-point
    n_20       10^20    * 1.40E+00
    T          eV       * 4.00E+01
    T_i/T_e             * 1.00E+00
    Delta_x    m          1.14E-02
    rho*                  2.03E-02
    nu*                   2.13E+01
    beta                  2.80E-04
    D/l_cold              2.27E+00
    D/l_hot               6.07E-01
    D/rho_p               8.35E+00
|medskip
    Target
    n_20       10^20    * 2.44E+00
    T          eV       * 5.20E+00
    T_i/T_e             * 1.00E+00
    Delta_t    m          4.32E-03
    rho*                  1.94E-02
    nu*                   2.19E+03
    beta                  6.34E-05
    D/l_cold              7.64E-01
    D/l_hot               5.66E-01
    D/rho_p               8.77E+00
    alpha_t               3.01E-02
    Denotes input value *

|endvb

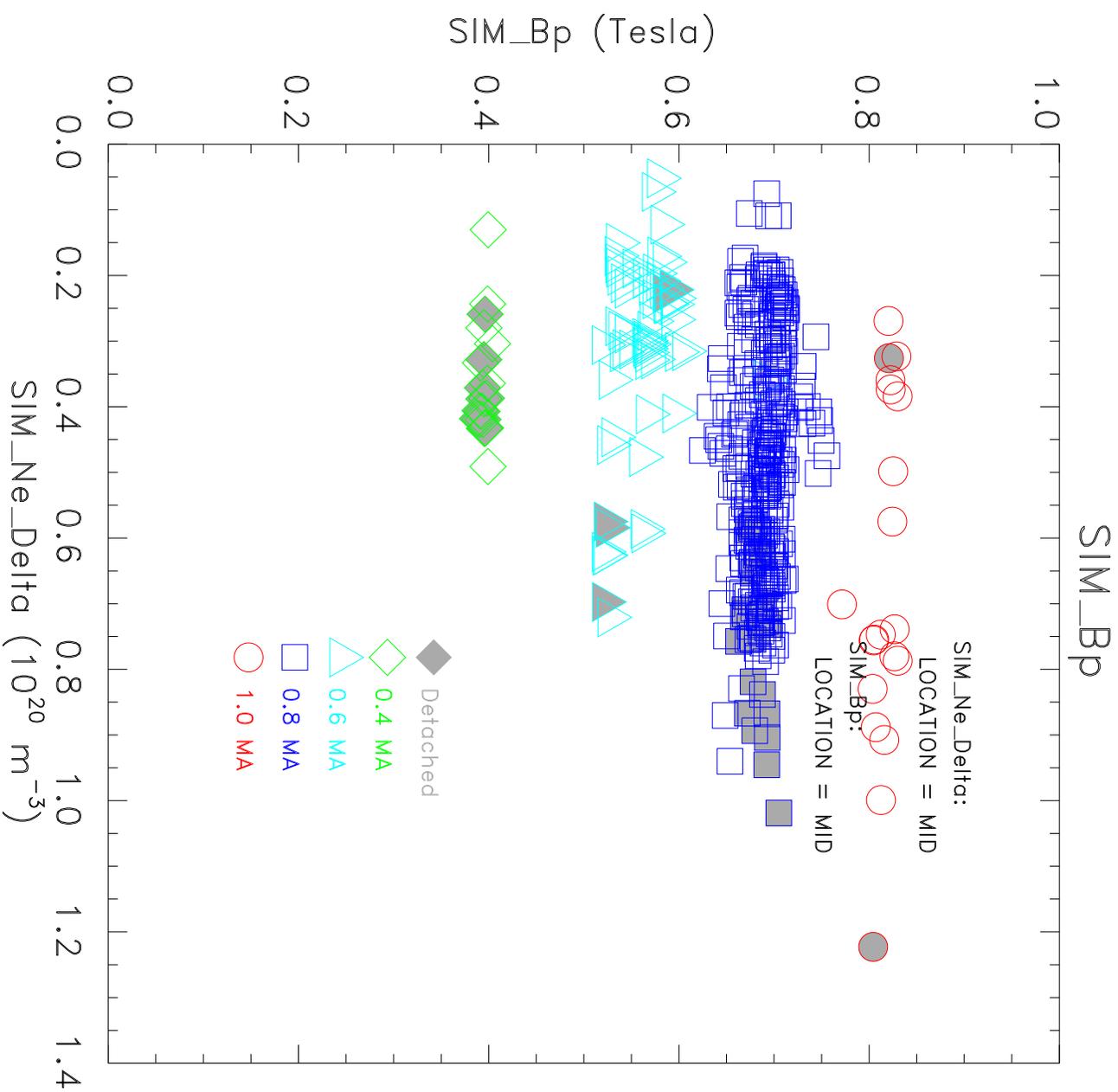
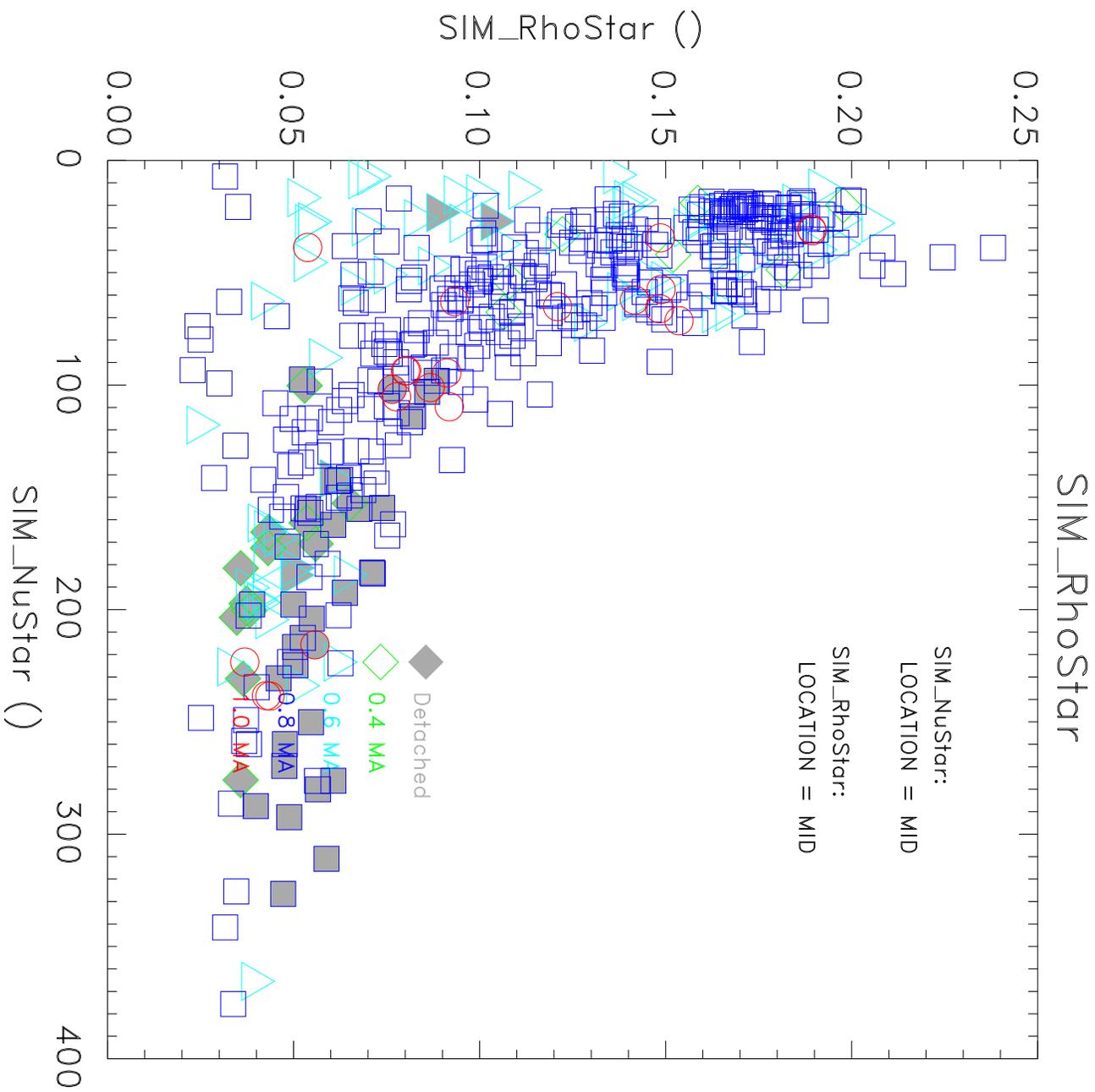
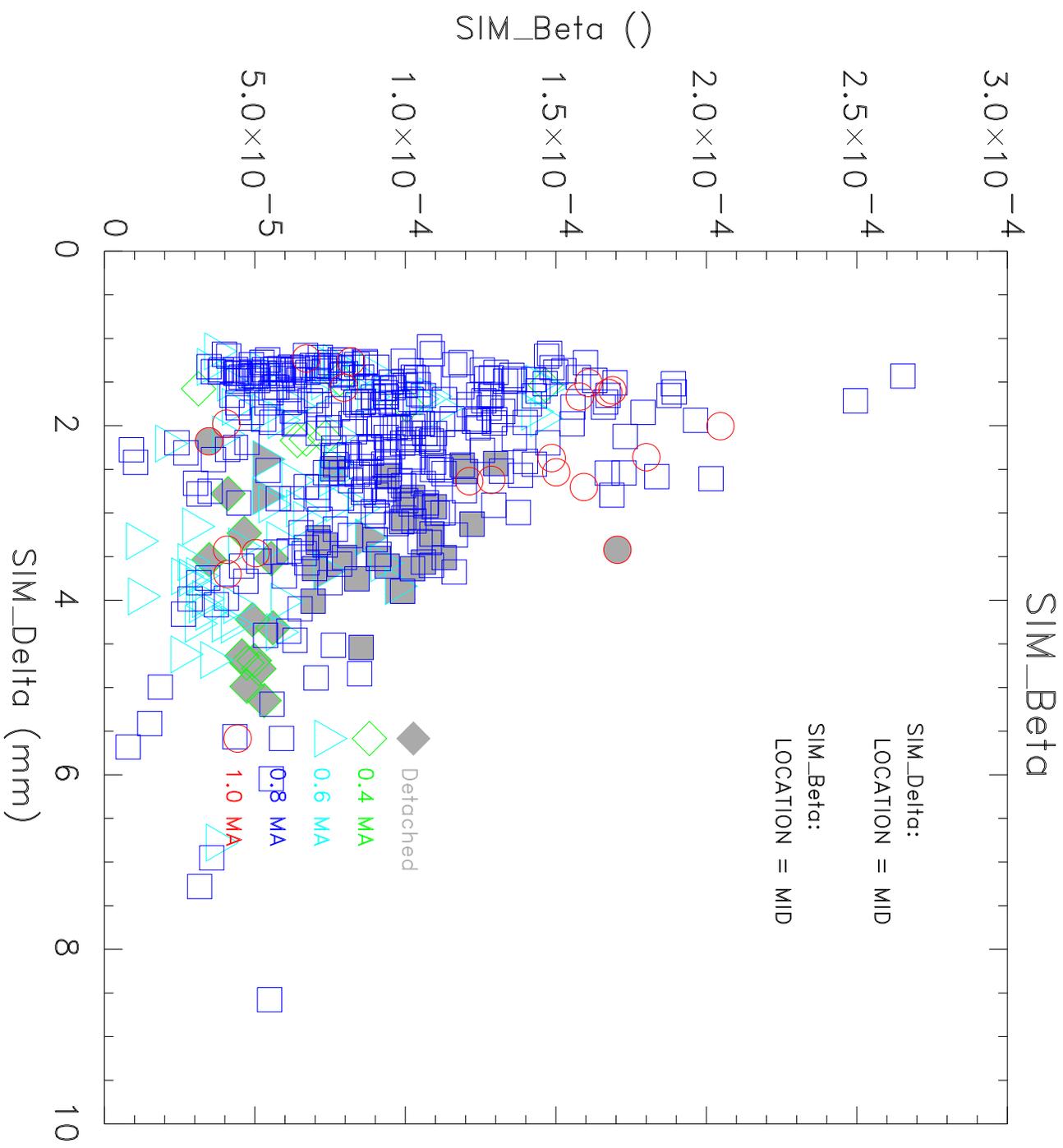
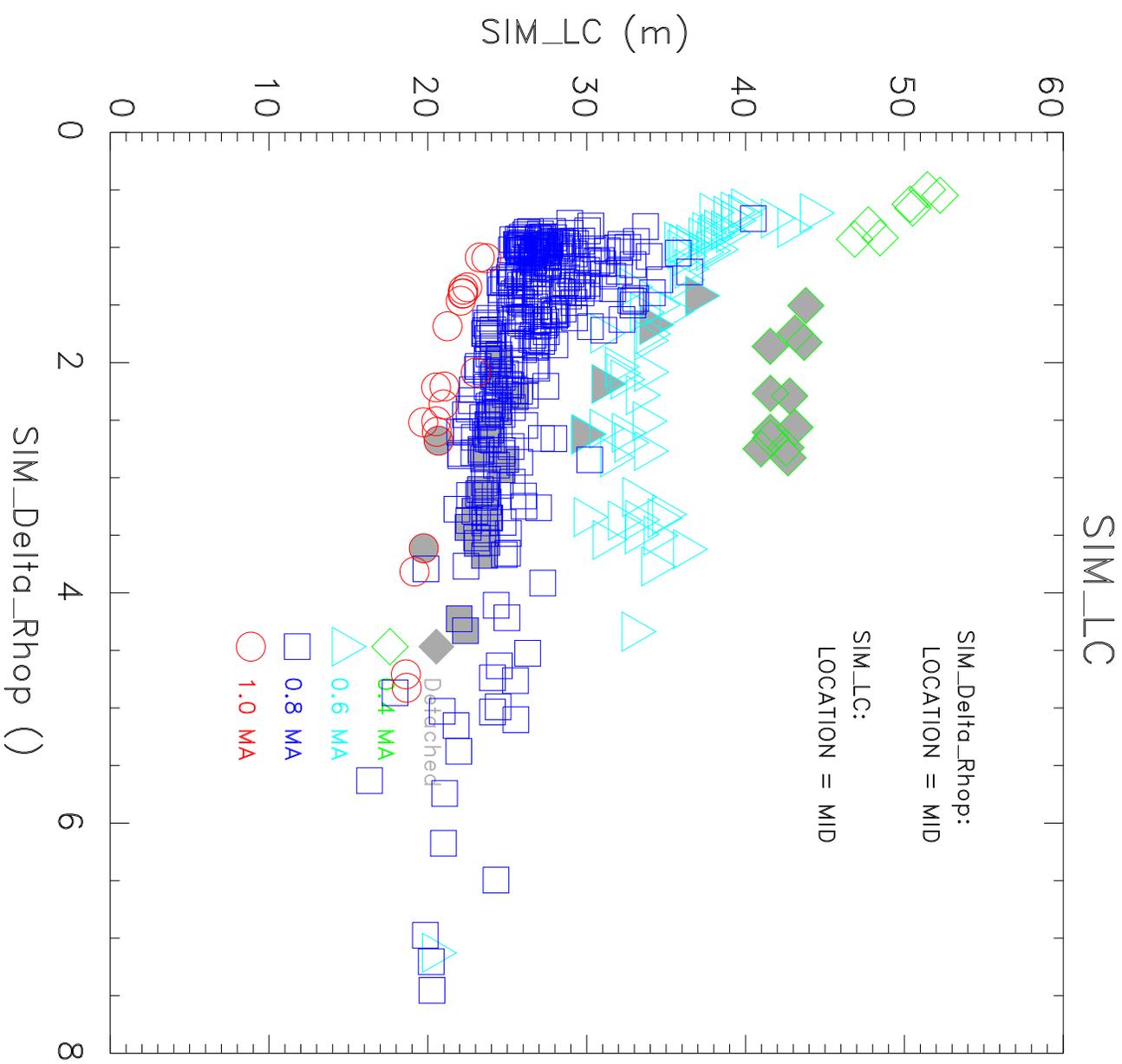
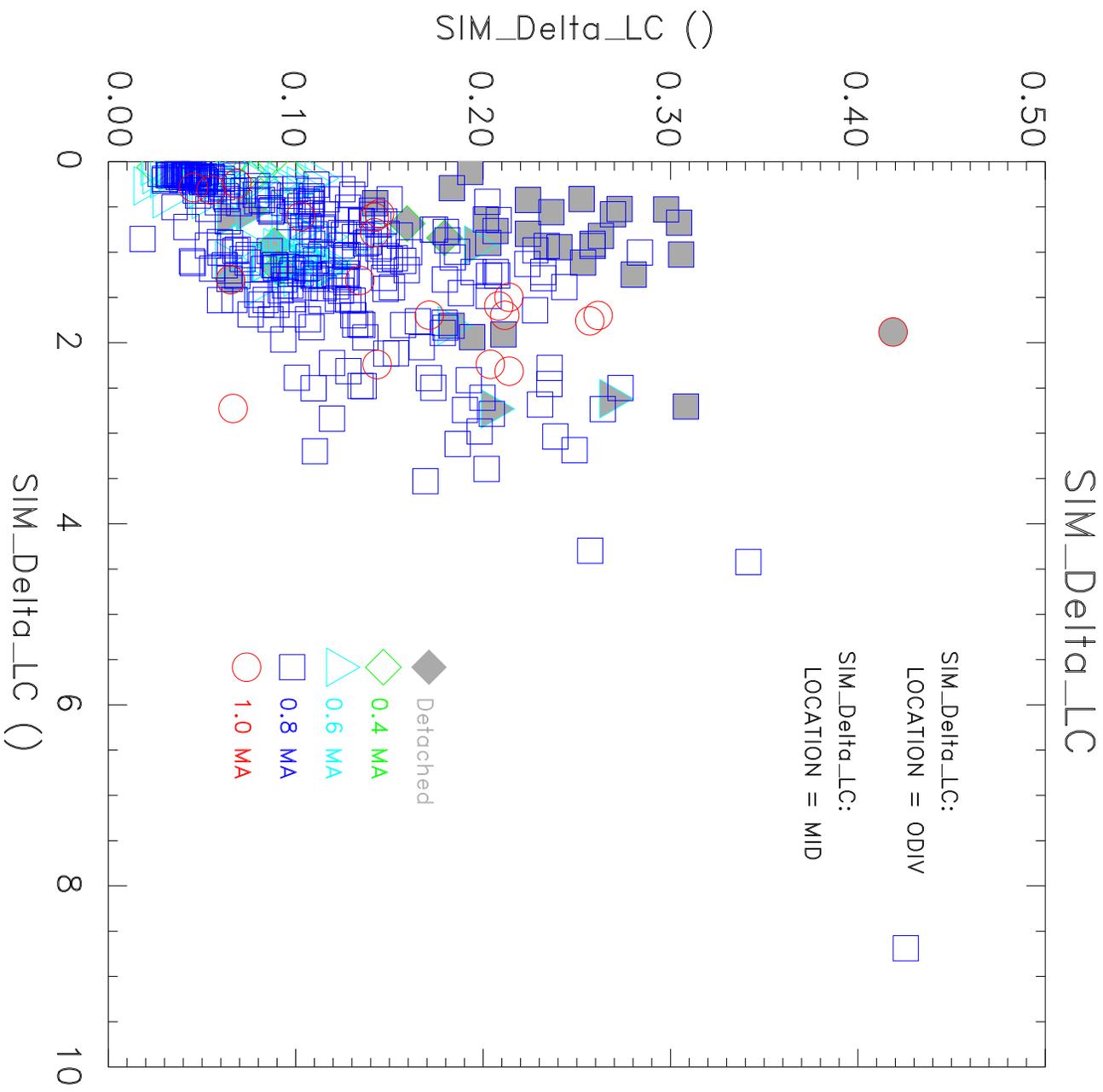
\end